\documentclass{article} 
\usepackage{iclr2022_conference,times}

\usepackage{hyperref}
\usepackage{url}
\usepackage[T1]{fontenc}

\title{Considerations for Multilingual Wikipedia Research}

\author{Isaac Johnson \& Emily Lescak  \\
Wikimedia Foundation \\
\texttt{\{isaac,elescak\}@wikimedia.org}}

\iclrfinalcopy 
\begin{document}

\maketitle

\begin{abstract}
English Wikipedia has long been an important data source for much research and natural language machine learning modeling. The growth of non-English language editions of Wikipedia, greater computational resources, and calls for equity in the performance of language and multimodal models have led to the inclusion of many more language editions of Wikipedia in datasets and models. Building better multilingual and multimodal models requires more than just access to expanded datasets; it also requires a better understanding of what is in the data and how this content was generated. This paper seeks to provide some background to help researchers think about what differences might arise between different language editions of Wikipedia and how that might affect their models. It details three major ways in which content differences between language editions arise (local context, community and governance, and technology) and recommendations for good practices when using multilingual and multimodal data for research and modeling.
\end{abstract}

\section{Introduction}
Wikipedia has been around for over 20 years and has been studied and modeled by researchers for almost as many, but it is not a monolithic resource. By 2005, there were already almost 200 language editions of Wikipedia and, as of 2022, there are over 300.\footnote{\url{https://meta.wikimedia.org/wiki/List_of_Wikipedias}} \cite{voss2005measuring} analyzed the structure and differences between several language editions (German, Japanese, Danish, and Croatian) as early as 2005, notably excluding English Wikipedia due to 36,000 geographic articles that were bot-generated, which ``biases some statistics.'' Since then, however, the focus has largely been on the textual content of English Wikipedia. It is only recently that the research community has expanded their work to be more multilingual---including more languages, especially non-English and smaller language editions---and multimodal---also studying the images and other multimedia content within Wikipedia and corresponding media repository Wikimedia Commons.

While multilingual and multimodal models are an important step forward, it is likewise important that researchers understand their increasingly multifaceted Wikipedia datasets so that they may use them responsibly. It is likely not necessary to exclude full language editions as \cite{voss2005measuring} did, but the dynamics that he described (bot-generated content and the skew that that can introduce as well as other aspects described in this work) are still present in many language editions of Wikipedia.\footnote{In fact, many of those 36,000 bot-generated articles mentioned by Voss about towns and cities on English Wikipedia remain largely unchanged per \cite{johnson2016not}.}

This paper seeks to support researchers working with multilingual and multimodal Wikipedia content through a survey of findings about the differences in content across languages and how they arise. These findings from the literature are complemented with the authors' own personal experiences when working with Wikimedia data and conversations with other researchers. Three major sources of differences are identified and detailed with examples: the local context, editor communities and governance, and technologies used by editors. Many of these facets are relevant even when working with a single language edition of Wikipedia, but their relevance becomes more pronounced when trying to handle multiple language editions in a consistent manner. The paper finishes with a few recommendations for good practices when doing this research and open areas of future research.

\section{Background}
We provide context on the role of Wikipedia data in large language models, a brief summary of the importance of understanding data quality, and some of the research that has more deeply explored differences across languages in Wikipedia.

\subsection{Wikipedia in Large Language Models}
\label{sec:wikiprocessing}
Many (perhaps all) large language models have made extensive use of text from Wikipedia to help train their models. Wikipedia is a substantive component of The Pile (\cite{gaopile}) and Common Crawl snapshots (\cite{dodge2021documenting}) that are used in training models like Megatron-Turing NLG 530B (\cite{smith2022using}), more than half of the data used to train the original BERT model (\cite{devlin2018bert}), and was explicitly excluded from the WebText corpora used in training the GPT-2 and GPT-3 models due to its appearance in the other datasets that were used (\cite{radford2019language}). It also figures prominently in image-text datasets such as the Wikipedia-Based Image Text (WIT) Dataset (\cite{srinivasan2021wit}) and LAION-400M (\cite{schuhmann2021laion}), the latter of which is derived from the Common Crawl and thus almost certainly includes Wikipedia articles.\footnote{A quick examination of a sample of the data confirmed this fact.} Wikipedia also is an important source of many standard benchmark datasets -- e.g., Wikitext for language modeling (\cite{merity2016pointer}) and SQuAD for reading comprehension (\cite{rajpurkar2016squad}).

For the preprocessing of Wikipedia datasets, there are two high-level questions that must be asked: 1) what articles (or images) are included, and, 2) for articles, what text cleaning steps are used to convert the raw markdown used to write Wikipedia articles (wikitext) into natural language free of extraneous syntax or noisy text. Regarding the first question, almost all Wikipedia datasets for training language models use all articles in the main namespace (regardless of quality and subject). Accordingly, image datasets such as WIT often use images that are included within Wikipedia articles (\cite{srinivasan2021wit}). Wikipedia-derived benchmarks, however, such as Wikitext and SQuAD do filter the articles extensively and often only use high-quality articles (featured articles and articles with high PageRank values respectively). An exception to this is \cite{guo2020wiki}, who released a pre-processed multilingual Wikipedia dataset for use in language models and took the additional step of filtering out pages unlikely to have high-quality text such as disambiguation pages and list articles. The preprocessing pipelines for parsing wikitext also vary greatly -- e.g., a simple Perl script used by fastText that removes some basic wikitext syntax\footnote{\url{https://github.com/facebookresearch/fastText/blob/main/wikifil.pl}}, a more fully-fledged Python tool used by BERT that removes some syntax but also expands templates on Wikipedia\footnote{\url{https://github.com/attardi/wikiextractor}}, a Python script used by HuggingFace that has its own process for cleaning syntax,\footnote{\url{https://github.com/huggingface/datasets/blob/master/datasets/wikipedia/wikipedia.py}} and the process described in \cite{guo2020wiki} that indicates some of the syntax/objects that it seeks to remove like tables, lists, and reference sections that are unlikely to have high-quality text (though no code is provided).

\subsection{Importance of data to modeling}
While datasets are highly important to the functioning of models, in-depth studies of their quality and approaches to improving that quality are often undervalued (\cite{gebru2021datasheets,sambasivan2021everyone,paullada2021data}). For instance, studies that have examined some of the multilingual and multimodal large-scale web-crawled datasets (that often contain Wikipedia as detailed above) find many issues. \cite{kreutzer2021quality} provides an in-depth look into the quality of several large-scale multilingual datasets and found substantive errors, including in the parallel sentence dataset WikiMatrix (\cite{schwenk2021wikimatrix}) that is extracted from Wikipedia articles. For WikiMatrix, the issues seemed to arise due to the automated process used to find parallel sentences but also stemmed from some of the ways in which Wikipedia language editions are divided up. \cite{kreutzer2021quality} goes on to describe some of the potential negative downstream effects of these issues such as reduced model quality, representation washing (false inclusion of low-resourced languages), and trust in incorrect ``facts'' extracted from models due to improper alignment of sentences. To quote them, ``data cleaning is no trivial task!''. Quality issues and concerns extend to multimodal datasets as well. \cite{birhane2021multimodal} examine the images and captions in LAION-400M (\cite{schuhmann2021laion}) and found many examples of disturbing imagery and highly-problematic stereotypes or captions.

\subsection{Multilingual Wikipedia research}
Much research (a lot of which referenced below comes from the field of computational social science) has examined differences between different language editions of Wikipedia from the standpoints of content (e.g., \cite{bao2012omnipedia,warncke2012search,he2018the_tower_of_babel,beytia2021visual}), readers (e.g., \cite{johnson2021global,lemmerich2019world,arora2022wikipedia}), and editors (e.g., \cite{kim2016understanding,bipat2018we,sen2015barriers}).\footnote{For a fuller accounting of the different gaps across the Wikimedia projects, see \cite{redi2020taxonomy}} These multilingual and multimodal Wikipedia studies provide insight into what differences arise across language editions and some of the factors that may cause these differences (further detailed in Section~\ref{sec:factors}). Much of these differences relate back to coverage and quality of text in Wikipedia and thus the quality of machine learning models trained on this content. While some coverage biases on Wikipedia are relatively consistent across language editions such as the gender gap,\footnote{\url{https://humaniki.wmcloud.org/}} others vary greatly based on what places and cultures are written about in a given language (\cite{miquel2019wikipedia}). Better understanding these differences can help researchers and practitioners to construct higher-quality datasets from Wikipedia and understand what is and is not included.

\section{Factors affecting Wikipedia data across languages}
\label{sec:factors}
Below, we outline some of the sources of major differences in content (and therefore data) between Wikipedia language editions based on an informal survey of research and documentation about the Wikipedia projects. These facets are certainly incomplete but cover a wide range of aspects to be aware of when working with multilingual and multimodal Wikipedia content. The impacts of some of these facets are direct---e.g., access to translation tools that greatly affect what content is created (articles or images that already exist on other language editions, often English, as described in \cite{warncke2012search}) and the style of writing (the content often starts from machine translation)---but others are far more subtle though no less important---e.g., barriers to new editors that might reduce the diversity of an editor community and therefore its content.

\subsection{Local context}
\label{sec:fac-context}
The first set of facets touches on the offline context for editors of a particular language edition.

\subsubsection{Geography and culture}
The Wikipedia projects are divided up by language and not geography. This simple fact has two major implications.

First, the different language editions of Wikipedia serve very different communities (\cite{johnson2021global,lemmerich2019world}) and thus often cover very different topics (\cite{bao2012omnipedia,hecht2009measuring,miquel2020wikipedia}). The resulting variation in quality and quantity of content about different topics (\cite{lewoniewski2017relative}) presumably would affect the resulting vocabulary and ability of language models trained on Wikipedia to accurately handle different topics. Even when language editions cover the same topics, editors often choose culturally-relevant imagery to include (\cite{he2018the_tower_of_babel}).

Second, not only are different language editions contextualized within different cultures (\cite{chelsy2019detecting}), but the diversity of these cultures varies greatly as well. On one end are languages like English, French, or Arabic that have been associated with colonialism or large diasporas and thus are home to editors from many different countries and social contexts. On the other end are languages like Croatian or Japanese that are much more specific to a single country and social context. The position on this spectrum can have trivial effects on the content---e.g., variation in dialects or spellings of words such as American English and British English\footnote{\url{https://en.wikipedia.org/wiki/Wikipedia:Manual_of_Style\#National_varieties_of_English}} that could affect language models---as well as much larger impacts on the content. For an example of the latter, community diversity appears to be an important factor in creating high-quality content (\cite{shi2019wisdom}) maintaining Wikipedia's core content policy of Neutral Point of View\footnote{\url{https://en.wikipedia.org/wiki/Wikipedia:Neutral_point_of_view}}, which if not followed, can lead to highly problematic content, especially around history or political topics (e.g., \cite{croatianreport,Sato_2021}).

\subsubsection{Source availability}
As a tertiary source that itself depends on reliable sources to support what is written in it,\footnote{\url{https://en.wikipedia.org/wiki/Wikipedia:No_original_research}} Wikipedia reflects the world and its biases (\cite{Wikipedia_mirror_2018}). While this helps maintain the high quality and verifiability of content, it also introduces biases based on what topics or people are themselves well-documented. For languages with fewer resources, especially when they are not digital or accessible (\cite{Gill_2021}), writing high-quality content can be much more difficult and the perspectives presented can be drawn from non-local viewpoints based on the sources available (\cite{sen2015barriers,lewoniewski2017analysis}).

\subsection{Editor community and governance}
\label{sec:fac-community}
The second set of facets focuses on aspects of governance that affect what content is contributed and by whom to a given Wikipedia.

\subsubsection{Rules}
\label{sec:fac-com-rules}
The most concrete aspect of governance are the rules that are hard-coded into different Wikipedia language editions and govern who can contribute and how. These rules range from governing who can edit articles---e.g., while no accounts are required to edit content on most Wikipedia language editions, Portuguese Wikipedia requires an account to edit (\cite{ptwiki2021}), which is known to discourage new editors from contributing (\cite{valueipediting})---to whether articles can easily be generated based on translations of other language editions---e.g., an option only available for a select few editors on English Wikipedia\footnote{\url{https://en.wikipedia.org/wiki/Wikipedia:Content_translation_tool\#English_Wikipedia_restrictions}} but that is widely available on many smaller language editions (\cite{Ozurumba_2021})---to whether the Wikipedians who are able to do many core moderation tasks are native speakers of the language.\footnote{\url{https://meta.wikimedia.org/wiki/CheckUser_policy}}.

\subsubsection{Policies}
Along with rules, Wikipedia language editions also have core policies that guide what content can be included in that language edition. While the core content policies\footnote{Neutral point of view, verifiability, no original research: \url{https://en.wikipedia.org/wiki/Wikipedia:Core_content_policies}} are common across language editions, other policies can differ substantially (\cite{ucocresearch}). For example, English Wikipedia accepts fair-use imagery\footnote{\url{https://en.wikipedia.org/wiki/Wikipedia:Non-free_content}} that is not allowed in Wikimedia Commons (the main image repository for Wikipedia). As a result, English Wikipedia contains many images of company logos, movie posters, and other otherwise protected content that would not be found on other language editions and would greatly affect the types of media and captions in datasets built from English Wikipedia as opposed to Wikimedia Commons. What constitutes a reliable source also can vary greatly by language edition (\cite{unreliableguidelines}), which has massive implications for what topics are notable enough for inclusion and what points-of-view can be expressed in the content.

\subsubsection{Norms}
Even with similar rules and policies, different language editions likely will have different norms about how content is created. How work is coordinated---e.g., discussions on talk pages or Wikiprojects as opposed to off-wiki groups on Facebook, email listservs, or chat apps like Telegram---directly impacts any conversational data extracted from Wikipedia (e.g., \cite{hua2018wikiconv}) and metrics related to collaboration\footnote{For example, article depth: \url{https://meta.wikimedia.org/wiki/Wikipedia_article_depth}} but also might impact the resulting quality of articles (\cite{morgan2013project}). How articles are improved---e.g., in draft workspaces, through fewer but larger edits, through many incremental edits---would affect edit datasets and tools such as vandalism detection that aim to evaluate edits as well as datasets that aim to capture cleaned-up versions of articles created via translation.\footnote{\url{https://www.mediawiki.org/wiki/Content_translation/Published_translations}}

\subsubsection{Size, age, and composition}
All three prior facets (rules, policies, and norms) are further mediated by the size, age, and composition of a given Wikipedia project. Smaller and younger language editions generally lack tools to support their work\footnote{\url{https://meta.wikimedia.org/wiki/Small_wiki_toolkits}}. They also can be less of a target of vandalism and less entrenched in their norms (\cite{teblunthuis2018revisiting}) but might display more idiosyncracies in terms of their content due to more relaxed policies (\cite{keegan2017evolution}) or based on early editors' interests and abilities to generate content via bots or semi-automated means (\cite{Guldbrandsson_2013}). While there are large gaps in the diversity of all Wikipedia editor communities (\cite{redi2020taxonomy}), smaller communities almost by definition will have less diversity.

\subsection{Technologies}
\label{sec:fac-technology}
The third set of facets deals with the technologies used by Wikipedians to contribute to Wikipedia, which can vary substantially by language edition.

\subsubsection{Interfaces and tools}
While every language edition uses the same core MediaWiki software to edit their version of Wikipedia, much of the tooling and interfaces used by editors can be configured in various ways or exists outside of that core software in extensions (\cite{geiger2014bots}). A salient example of this is an extension called Structured Discussions\footnote{Structured Discussions (\url{https://www.mediawiki.org/wiki/Structured_Discussions}) has been more recently superseded by a more central talk pages project (\url{https://www.mediawiki.org/wiki/Talk_pages_project}).} (or ``Flow'') that, as the name suggests, provides more structure around talk page discussions. The increase in structure around conversations can have a substantive impact on the nature of discussions (\cite{aragon2017thread}). Other major extensions with mixed uptake include Flagged Revisions\footnote{\url{https://en.wikipedia.org/wiki/Wikipedia:Flagged_revisions}} (determines whether a revision is shown to a reader before review), Content Translation (discussed in Section~\ref{sec:fac-com-rules}), Newcomer Tasks\footnote{\url{https://www.mediawiki.org/wiki/Growth/Feature_summary}} (edit recommendations for new editors that include tasks such as adding images to articles) and PageAssessments\footnote{\url{https://www.mediawiki.org/wiki/Extension:PageAssessments}} (simplifies tagging of content by Wikiprojects with topic, quality, and importance annotations).

\subsubsection{Bots and filters}
Bots and filters are also examples of tools that vary widely and have an even more direct relationship to edits on Wikipedia. Taking the example of vandalism detection, English Wikipedia has an extensive AbuseFilter configuration\footnote{\url{https://en.wikipedia.org/wiki/Wikipedia:Edit_filter}} that catches many bad edits before they are even published, automated bots such as ClueBot~NG\footnote{\url{https://en.wikipedia.org/wiki/User:ClueBot_NG}}, and RecentChanges filters\footnote{\url{https://www.mediawiki.org/wiki/ORES/RCFilters}} based on machine learning models for detecting bad-faith and damaging edits (\cite{halfaker2020ores}). This suite of tooling both reduces the amount of vandalism that reaches Wikipedia and greatly speeds up the response time of editors to vandalism by helping them to quickly identify problematic edits (\cite{geiger2013levee}). English Wikipedia, as the largest language community, has many specialized tools that blur the lines between human and bot editing, such as AutoWikiBrowser or Twinkle\footnote{\url{https://en.wikipedia.org/wiki/Wikipedia:AutoWikiBrowser} and \url{https://en.wikipedia.org/wiki/Wikipedia:Twinkle} respectively. For more tools, see \url{https://en.wikipedia.org/wiki/Special:Tags}} and, as a result, has a very high rate of small edits done through these tools as a part of routine maintenance tasks.

Awareness of the origin of content and edits can help greatly in identifying what might be more natural language as opposed to large-scale, rule-based language generation. While bot-based generation of entire articles is less common on English Wikipedia now, many articles still have their origins as bot-generated and have been minimally edited since then (\cite{johnson2016not}). This content is largely standard text with the key facts differing between articles---e.g., the population of a town or its geographic neighbors, all of which can be automatically extracted from databases. Tool-based editing to, for example, fix common misspellings, could skew grammatical error correction corpora built from Wikipedia (e.g., \cite{lichtarge2019corpora}). Tools exist to identify these patterns (e.g., \cite{alshomary2019wikipedia}) and a high-quality language model might seek to deduplicate this content or downweight it in training.

\subsubsection{Transcluded content}
\cite{mitrevski2020wikihist} and \cite{johnson2020analyzing} have documented a growing trend on certain Wikipedia language editions in the shift from content being written within the wikitext of an article to content being transcluded from other pages or via complicated logic---e.g., copying a standard table of links for all articles about a particular region, auto-filling infoboxes with standard facts such as a town's population or a country's flag. This brings consistency for readers and eases the maintenance burden for editors (only have to update content in one place, not in all the articles in which it is used), but it also leads to a growing disconnect between the content directly expressed in an article's wikitext and the parsed content that comprises the article. As with the discussion of bots and filters above, filtering out this highly-structured, templated content may be desirable for language models in some circumstances. However, for research that relies on extracting facts from infoboxes or using links to build a graph representation of Wikipedia, a large difference would be seen between the data extracted from wikitext and the data extracted from the final HTML of a page.\footnote{The final HTML of articles had long only been available via scraping or APIs, but complete HTML dumps for Wikipedia are now available to researchers: \url{https://dumps.wikimedia.org/other/enterprise_html/}.}

\section{Recommendations}
\label{sec:recommendations}
The above sections identified reasons why the content in different Wikipedia language editions can vary in quality, topic, and appropriateness for various multilingual and multimodal tasks. Below, we offer recommendations for steps to address some of these concerns and better tune Wikipedia-based datasets to a particular modeling task.

\subsection{Situated researchers}
\label{sec:situated}
The first and perhaps most important recommendation is to have researchers\footnote{Using this term broadly to not just mean academics but anyone involved in understanding and analyzing the content or data} who are situated within the communities being studied. This is not specific to Wikipedia research, with calls for this level of inclusion and participatory research such as for natural language processing (NLP) research more broadly (\cite{nekoto2020participatory}) and data governance (\cite{carroll19care}). Having a research team that includes Wikipedians---i.e. editors---from that language community, native speakers, and folks who are long-time readers of a language edition is an important step beyond e.g., using machine-translated versions of content for inspecting your data. These situated researchers (e.g., \cite{hickman2021understanding,bipat2021wikipedia,unreliableguidelines}) can help the broader team to understand nuances in the language and identify factors like the local context, governance, and technologies above.

For researchers without existing ties to Wikimedians, there are still many ways to connect with the editor communities. While care should be taken to respect Wikipedia (as with research involving any online community),\footnote{\url{https://en.wikipedia.org/wiki/Wikipedia:What_Wikipedia_is_not\#Wikipedia_is_not_a_laboratory}} research with clear benefits to Wikipedia is generally welcomed. Beyond individual Wikimedians, there are also more formal organizations of Wikimedians ranging from very simple user groups to the more organized chapters\footnote{\url{https://meta.wikimedia.org/wiki/Wikimedia_movement_affiliates}} to the Wikimedia Foundation.\footnote{\url{https://wikimediafoundation.org/}} As laid out by \cite{voss2005measuring}, there are many convenings\footnote{\url{https://meta.wikimedia.org/wiki/Events}} for sharing work and meeting editors, organizers, and other folks connected to the Wikimedia movement. There are also several forms of support that are available to researchers including mailing lists,\footnote{\url{https://lists.wikimedia.org/postorius/lists/wiki-research-l.lists.wikimedia.org/}} numerous research venues such as the long-running WikiWorkshop series,\footnote{\url{https://wikiworkshop.org}} and technical resources to support tool-building or Wikimedia research.\footnote{\url{https://wikitech.wikimedia.org/wiki/Help:Cloud_Services_Introduction}} 

\subsection{Language-agnostic metrics}
Language model performance is strongly tied to the amount of data available for that language---e.g., \cite{wu2020all}'s analysis of multilingual BERT on many low-resourced languages. When models are trained for Wikipedia-specific tasks, however, there are many language-agnostic features that can potentially be used to boost performance in lower-resourced languages. For example, in training topic models for Wikipedia, \cite{piccardi2021crosslingual} and \cite{johnson2021language} relied not on the words in an article but the links. Article links can be mapped to language-agnostic Wikidata IDs such that e.g., a link on English Wikipedia to the article for poblano peppers will be represented identically as a link to chiles poblanos\footnote{Q897746: \url{https://www.wikidata.org/wiki/Q897746}}. This shared vocabulary means that lower-resourced languages can benefit from training data in better-resourced languages as long as they have corresponding articles and greatly reduces the amount of language-specific data needed. Other language-agnostic approaches include graph-based modeling (using links and Wikidata IDs as described for topic modeling) or taking advantage of other semi-structured components of articles---e.g., templates, sections, categories, Wikidata statements---to generate features that can be represented in a language-agnostic manner (e.g., \cite{lewoniewski2017relative,beytia2021visual}).

\subsection{Matched corpora}
As detailed above, there are many outliers that can skew datasets generated from Wikipedia. Sometimes these outliers are of interest but oftentimes they can just add noise or otherwise obscure what the researcher would like to model. Building matched corpora can help reduce the impact of this skew in training or evaluating models. For example, \cite{field2020controlled} studied the relationship between an individual's identity and how they were written about on Wikipedia. They built a matched corpora in which each article about a woman was matched with a similar article about a man (based on article categories) and then computed their metrics of interest. This matched corpora was important to isolating the specific effect of an individual's gender while holding constant their notability or field of study. For multilingual analyses, an obvious matched corpora is one in which the same articles are included across language editions. While this could entail a large loss of data, particularly around locally-important topics that might not appear in other language editions, it would bring important consistency to e.g., benchmark datasets. Care must be taken when building matched datasets that are more fine-grained than article-level, as e.g., with the parallel sentence dataset WikiMatrix as described in \cite{kreutzer2021quality}. 

\subsection{Future research and opportunities}
While this work seeks to lay out some factors to be aware of when working with Wikipedia data and good practices to follow, it cannot yet provide best practices. Future work and convenings such as Wiki-M3L\footnote{\url{https://meta.wikimedia.org/wiki/Wiki-M3L}} or the Wikimedia community events mentioned in Section~\ref{sec:situated} will be necessary to build these tools and norms. As identified in Section~\ref{sec:wikiprocessing}, a good starting place would be shared code and parameterizations for effectively preprocessing Wikipedia text. These practices should be supported by robust methodological research like \cite{mitrevski2020wikihist} (who examined the difference between wikitext and parsed HTML of Wikipedia articles) or \cite{hill2014consider} (who examined how including Wikipedia redirect pages affects studies of readership).

Researchers should also consider how their work can contribute back to the Wikimedia communities that have generated so much data that has supported progress in many fields of research. Challenges like the Wikipedia Image/Caption Matching content\footnote{\url{https://ai.googleblog.com/2021/09/announcing-wit-wikipedia-based-image.html}} are a good way to participate in research with clear benefits for the Wikimedia community. Machine translation researchers might find opportunities to expand the models available to Wikipedians for translating articles into their language.\footnote{\url{https://www.mediawiki.org/wiki/Content_translation\#How_to_participate}} In general, building connections with the Wikimedia community as described in Section~\ref{sec:situated} will help to identify opportunities for technical contributions to the Wikimedia projects.

\subsubsection*{Acknowledgments}
We would like to thank Taryn Bipat and Jacob Thebault-Spieker, whose research and presentations at the Wikimedia Research Showcase motivated this work. We also would like to thank the Wikimedia Foundation Research Team for their input and discussions as well as the many researchers and Wikimedians whose work and discussions we are building on.

\bibliography{iclr2022_conference}

\begin{thebibliography}{67}
\providecommand{\natexlab}[1]{#1}
\providecommand{\url}[1]{\texttt{#1}}
\expandafter\ifx\csname urlstyle\endcsname\relax
  \providecommand{\doi}[1]{doi: #1}\else
  \providecommand{\doi}{doi: \begingroup \urlstyle{rm}\Url}\fi

\bibitem[Alshomary et~al.(2019)Alshomary, V{\"o}lske, Licht, Wachsmuth, Stein,
  Hagen, and Potthast]{alshomary2019wikipedia}
Milad Alshomary, Michael V{\"o}lske, Tristan Licht, Henning Wachsmuth, Benno
  Stein, Matthias Hagen, and Martin Potthast.
\newblock Wikipedia text reuse: Within and without.
\newblock In \emph{European Conference on Information Retrieval}, pp.\
  747--754. Springer, 2019.

\bibitem[Arag{\'o}n et~al.(2017)Arag{\'o}n, G{\'o}mez, and
  Kaltenbrunner]{aragon2017thread}
Pablo Arag{\'o}n, Vicen{\c{c}} G{\'o}mez, and Andreaks Kaltenbrunner.
\newblock To thread or not to thread: The impact of conversation threading on
  online discussion.
\newblock In \emph{Proceedings of the International AAAI Conference on Web and
  Social Media}, volume~11, pp.\  12--21, 2017.

\bibitem[Arora et~al.(2022)Arora, Gerlach, Piccardi, Garc{\'\i}a-Dur{\'a}n, and
  West]{arora2022wikipedia}
Akhil Arora, Martin Gerlach, Tiziano Piccardi, Alberto Garc{\'\i}a-Dur{\'a}n,
  and Robert West.
\newblock Wikipedia reader navigation: When synthetic data is enough.
\newblock In \emph{Proceedings of the Fifteenth ACM International Conference on
  Web Search and Data Mining}, WSDM '22, pp.\  16–26, New York, NY, USA,
  2022. Association for Computing Machinery.
\newblock ISBN 9781450391320.
\newblock \doi{10.1145/3488560.3498496}.
\newblock URL \url{https://doi.org/10.1145/3488560.3498496}.

\bibitem[Bao et~al.(2012)Bao, Hecht, Carton, Quaderi, Horn, and
  Gergle]{bao2012omnipedia}
Patti Bao, Brent Hecht, Samuel Carton, Mahmood Quaderi, Michael Horn, and
  Darren Gergle.
\newblock Omnipedia: bridging the wikipedia language gap.
\newblock In \emph{Proceedings of the SIGCHI Conference on Human Factors in
  Computing Systems}, pp.\  1075--1084, 2012.

\bibitem[Berson et~al.(2021)Berson, Sengul-Jones, and
  Tamani]{unreliableguidelines}
Amber Berson, Monika Sengul-Jones, and Melissa Tamani.
\newblock Unreliable guidelines: Reliable sources and marginalized communities
  in french, english and spanish wikipedias, 2021.
\newblock URL
  \url{https://artandfeminism.org/initiatives/current-initiatives/reading-together/}.

\bibitem[Beyt{\'\i}a et~al.(2022)Beyt{\'\i}a, Agarwal, Redi, and
  Singh]{beytia2021visual}
Pablo Beyt{\'\i}a, Pushkal Agarwal, Miriam Redi, and Vivek~K Singh.
\newblock Visual gender biases in wikipedia: A systematic evaluation across the
  ten most spoken languages.
\newblock Sixteenth International AAAI Conference on Web and Social Media,
  2022.

\bibitem[Bipat et~al.(2018)Bipat, McDonald, and Zachry]{bipat2018we}
Taryn Bipat, David~W McDonald, and Mark Zachry.
\newblock Do we all talk before we type? understanding collaboration in
  wikipedia language editions.
\newblock In \emph{Proceedings of the 14th International Symposium on Open
  Collaboration}, pp.\  1--11, 2018.

\bibitem[Bipat et~al.(2021)Bipat, Alimohammadi, Yu, McDonald, and
  Zachry]{bipat2021wikipedia}
Taryn Bipat, Negin Alimohammadi, Yihan Yu, David~W McDonald, and Mark Zachry.
\newblock Wikipedia beyond the english language edition: How do editors
  collaborate in the farsi and chinese wikipedias?
\newblock \emph{Proceedings of the ACM on Human-Computer Interaction},
  5\penalty0 (CSCW1):\penalty0 1--39, 2021.

\bibitem[Birhane et~al.(2021)Birhane, Prabhu, and
  Kahembwe]{birhane2021multimodal}
Abeba Birhane, Vinay~Uday Prabhu, and Emmanuel Kahembwe.
\newblock Multimodal datasets: misogyny, pornography, and malignant
  stereotypes.
\newblock \emph{arXiv preprint arXiv:2110.01963}, 2021.

\bibitem[Carroll et~al.(2020)Carroll, Garba, Figueroa-Rodr{\'\i}guez, Holbrook,
  Lovett, Materechera, Parsons, Raseroka, Rodriguez-Lonebear, Rowe,
  et~al.]{carroll19care}
Stephanie~Russo Carroll, Ibrahim Garba, Oscar~L Figueroa-Rodr{\'\i}guez, Jarita
  Holbrook, Raymond Lovett, Simeon Materechera, Mark Parsons, Kay Raseroka,
  Desi Rodriguez-Lonebear, Robyn Rowe, et~al.
\newblock The care principles for indigenous data governance.
\newblock \emph{Data Science Journal 19 (1)}, 2020.

\bibitem[Chelsy~Xie et~al.(2019)Chelsy~Xie, Johnson, and
  Gomez]{chelsy2019detecting}
Xiaoxi Chelsy~Xie, Isaac Johnson, and Anne Gomez.
\newblock Detecting and gauging impact on wikipedia page views.
\newblock In \emph{Companion Proceedings of The 2019 World Wide Web
  Conference}, pp.\  1254--1261, 2019.

\bibitem[Devlin et~al.(2018)Devlin, Chang, Lee, and Toutanova]{devlin2018bert}
Jacob Devlin, Ming-Wei Chang, Kenton Lee, and Kristina Toutanova.
\newblock Bert: Pre-training of deep bidirectional transformers for language
  understanding.
\newblock \emph{arXiv preprint arXiv:1810.04805}, 2018.

\bibitem[Dodge et~al.(2021)Dodge, Sap, Marasovi{\'c}, Agnew, Ilharco,
  Groeneveld, Mitchell, and Gardner]{dodge2021documenting}
Jesse Dodge, Maarten Sap, Ana Marasovi{\'c}, William Agnew, Gabriel Ilharco,
  Dirk Groeneveld, Margaret Mitchell, and Matt Gardner.
\newblock Documenting large webtext corpora: A case study on the colossal clean
  crawled corpus.
\newblock In \emph{Proceedings of the 2021 Conference on Empirical Methods in
  Natural Language Processing}, pp.\  1286--1305, 2021.

\bibitem[Field et~al.(2020)Field, Park, and Tsvetkov]{field2020controlled}
Anjalie Field, Chan~Young Park, and Yulia Tsvetkov.
\newblock Controlled analyses of social biases in wikipedia bios.
\newblock \emph{arXiv preprint arXiv:2101.00078}, 2020.

\bibitem[Gao et~al.(2020)Gao, Biderman, Black, Golding, Hoppe, Foster, Phang,
  He, Thite, Nabeshima, et~al.]{gaopile}
Leo Gao, Stella Biderman, Sid Black, Laurence Golding, Travis Hoppe, Charles
  Foster, Jason Phang, Horace He, Anish Thite, Noa Nabeshima, et~al.
\newblock The pile: An 800gb dataset of diverse text for language modeling.
\newblock 2020.

\bibitem[Gebru et~al.(2021)Gebru, Morgenstern, Vecchione, Vaughan, Wallach,
  Iii, and Crawford]{gebru2021datasheets}
Timnit Gebru, Jamie Morgenstern, Briana Vecchione, Jennifer~Wortman Vaughan,
  Hanna Wallach, Hal~Daum{\'e} Iii, and Kate Crawford.
\newblock Datasheets for datasets.
\newblock \emph{Communications of the ACM}, 64\penalty0 (12):\penalty0 86--92,
  2021.

\bibitem[Geiger(2014)]{geiger2014bots}
R~Stuart Geiger.
\newblock Bots, bespoke, code and the materiality of software platforms.
\newblock \emph{Information, Communication \& Society}, 17\penalty0
  (3):\penalty0 342--356, 2014.

\bibitem[Geiger \& Halfaker(2013)Geiger and Halfaker]{geiger2013levee}
R~Stuart Geiger and Aaron Halfaker.
\newblock When the levee breaks: without bots, what happens to wikipedia's
  quality control processes?
\newblock In \emph{Proceedings of the 9th International Symposium on Open
  Collaboration}, pp.\  1--6, 2013.

\bibitem[Gill(2021)]{Gill_2021}
Satdeep Gill.
\newblock Balinese wikisource (wikipustaka) gives a new life to palm-leaf
  manuscripts, Nov 2021.
\newblock URL
  \url{https://diff.wikimedia.org/2021/11/15/balinese-wikisource-wikipustaka-gives-a-new-life-to-palm-leaf-manuscripts/}.

\bibitem[Guldbrandsson(2013)]{Guldbrandsson_2013}
Lennart Guldbrandsson.
\newblock Swedish wikipedia surpasses 1 million articles with aid of article
  creation bot, Jun 2013.
\newblock URL
  \url{https://diff.wikimedia.org/2013/06/17/swedish-wikipedia-1-million-articles/}.

\bibitem[Guo et~al.(2020)Guo, Dai, Vrande{\v{c}}i{\'c}, and
  Al-Rfou]{guo2020wiki}
Mandy Guo, Zihang Dai, Denny Vrande{\v{c}}i{\'c}, and Rami Al-Rfou.
\newblock Wiki-40b: Multilingual language model dataset.
\newblock In \emph{Proceedings of the 12th Language Resources and Evaluation
  Conference}, pp.\  2440--2452, 2020.

\bibitem[Halfaker \& Geiger(2020)Halfaker and Geiger]{halfaker2020ores}
Aaron Halfaker and R~Stuart Geiger.
\newblock Ores: Lowering barriers with participatory machine learning in
  wikipedia.
\newblock \emph{Proceedings of the ACM on Human-Computer Interaction},
  4\penalty0 (CSCW2):\penalty0 1--37, 2020.

\bibitem[He et~al.(2018)He, Lin, Adar, and Hecht]{he2018the_tower_of_babel}
Shiqing He, Allen~Yilun Lin, Eytan Adar, and Brent Hecht.
\newblock The\_tower\_of\_babel. jpg: diversity of visual encyclopedic
  knowledge across wikipedia language editions.
\newblock In \emph{Twelfth International AAAI Conference on Web and Social
  Media}, 2018.

\bibitem[Hecht \& Gergle(2009)Hecht and Gergle]{hecht2009measuring}
Brent Hecht and Darren Gergle.
\newblock Measuring self-focus bias in community-maintained knowledge
  repositories.
\newblock In \emph{Proceedings of the fourth international conference on
  communities and technologies}, pp.\  11--20, 2009.

\bibitem[Hickman et~al.(2021)Hickman, Pasad, Sanghavi, Thebault-Spieker, and
  Lee]{hickman2021understanding}
Molly~G Hickman, Viral Pasad, Harsh~Kamalesh Sanghavi, Jacob Thebault-Spieker,
  and Sang~Won Lee.
\newblock Understanding wikipedia practices through hindi, urdu, and english
  takes on an evolving regional conflict.
\newblock \emph{Proceedings of the ACM on Human-Computer Interaction},
  5\penalty0 (CSCW1):\penalty0 1--31, 2021.

\bibitem[Hill \& Shaw(2014)Hill and Shaw]{hill2014consider}
Benjamin~Mako Hill and Aaron Shaw.
\newblock Consider the redirect: A missing dimension of wikipedia research.
\newblock In \emph{Proceedings of The International Symposium on Open
  Collaboration}, pp.\  1--4, 2014.

\bibitem[Hua et~al.(2018)Hua, Danescu-Niculescu-Mizil, Taraborelli, Thain,
  Sorensen, and Dixon]{hua2018wikiconv}
Yiqing Hua, Cristian Danescu-Niculescu-Mizil, Dario Taraborelli, Nithum Thain,
  Jeffery Sorensen, and Lucas Dixon.
\newblock Wikiconv: A corpus of the complete conversational history of a large
  online collaborative community.
\newblock In \emph{Proceedings of the Conference on Empirical Methods in
  Natural Language Processing}, 2018.

\bibitem[Johnson(2020)]{johnson2020analyzing}
Isaac Johnson.
\newblock Analyzing wikidata transclusion on english wikipedia.
\newblock In \emph{Companion Proceedings of the 20th International Conference
  on Semantic Web}, 2020.

\bibitem[Johnson et~al.(2021{\natexlab{a}})Johnson, Gerlach, and
  S{\'a}ez-Trumper]{johnson2021language}
Isaac Johnson, Martin Gerlach, and Diego S{\'a}ez-Trumper.
\newblock Language-agnostic topic classification for wikipedia.
\newblock In \emph{Companion Proceedings of the Web Conference 2021}, pp.\
  594--601, 2021{\natexlab{a}}.

\bibitem[Johnson et~al.(2021{\natexlab{b}})Johnson, Lemmerich,
  S{\'a}ez-Trumper, West, Strohmaier, and Zia]{johnson2021global}
Isaac Johnson, Florian Lemmerich, Diego S{\'a}ez-Trumper, Robert West, Markus
  Strohmaier, and Leila Zia.
\newblock Global gender differences in wikipedia readership.
\newblock In \emph{Proceedings of the International AAAI Conference on Web and
  Social Media}, volume~15, pp.\  254--265, 2021{\natexlab{b}}.

\bibitem[Johnson et~al.(2016)Johnson, Lin, Li, Hall, Halfaker, Sch{\"o}ning,
  and Hecht]{johnson2016not}
Isaac~L Johnson, Yilun Lin, Toby Jia-Jun Li, Andrew Hall, Aaron Halfaker,
  Johannes Sch{\"o}ning, and Brent Hecht.
\newblock Not at home on the range: Peer production and the urban/rural divide.
\newblock In \emph{Proceedings of the 2016 CHI conference on Human Factors in
  Computing Systems}, pp.\  13--25, 2016.

\bibitem[Keegan \& Fiesler(2017)Keegan and Fiesler]{keegan2017evolution}
Brian Keegan and Casey Fiesler.
\newblock The evolution and consequences of peer producing wikipedia's rules.
\newblock In \emph{Proceedings of the International AAAI Conference on Web and
  Social Media}, volume~11, pp.\  112--121, 2017.

\bibitem[Kim et~al.(2016)Kim, Park, Hale, Kim, Byun, and
  Oh]{kim2016understanding}
Suin Kim, Sungjoon Park, Scott~A Hale, Sooyoung Kim, Jeongmin Byun, and Alice~H
  Oh.
\newblock Understanding editing behaviors in multilingual wikipedia.
\newblock \emph{PloS one}, 11\penalty0 (5):\penalty0 e0155305, 2016.

\bibitem[Kreutzer et~al.(2021)Kreutzer, Caswell, Wang, Wahab, van Esch,
  Ulzii-Orshikh, Tapo, Subramani, Sokolov, Sikasote,
  et~al.]{kreutzer2021quality}
Julia Kreutzer, Isaac Caswell, Lisa Wang, Ahsan Wahab, Daan van Esch,
  Nasanbayar Ulzii-Orshikh, Allahsera Tapo, Nishant Subramani, Artem Sokolov,
  Claytone Sikasote, et~al.
\newblock Quality at a glance: An audit of web-crawled multilingual datasets.
\newblock \emph{arXiv preprint arXiv:2103.12028}, 2021.

\bibitem[Lemmerich et~al.(2019)Lemmerich, S{\'a}ez-Trumper, West, and
  Zia]{lemmerich2019world}
Florian Lemmerich, Diego S{\'a}ez-Trumper, Robert West, and Leila Zia.
\newblock Why the world reads wikipedia: Beyond english speakers.
\newblock In \emph{Proceedings of the Twelfth ACM International Conference on
  Web Search and Data Mining}, pp.\  618--626, 2019.

\bibitem[Lewoniewski et~al.(2017{\natexlab{a}})Lewoniewski, W{\k{e}}cel, and
  Abramowicz]{lewoniewski2017analysis}
W{\l}odzimierz Lewoniewski, Krzysztof W{\k{e}}cel, and Witold Abramowicz.
\newblock Analysis of references across wikipedia languages.
\newblock In \emph{International Conference on Information and Software
  Technologies}, pp.\  561--573. Springer, 2017{\natexlab{a}}.

\bibitem[Lewoniewski et~al.(2017{\natexlab{b}})Lewoniewski, W\k{e}cel, and
  Abramowicz]{lewoniewski2017relative}
W{\l}odzimierz Lewoniewski, Krzysztof W\k{e}cel, and Witold Abramowicz.
\newblock Relative quality and popularity evaluation of multilingual wikipedia
  articles.
\newblock volume~4, pp.\ ~43. Multidisciplinary Digital Publishing Institute,
  2017{\natexlab{b}}.

\bibitem[Lichtarge et~al.(2019)Lichtarge, Alberti, Kumar, Shazeer, Parmar, and
  Tong]{lichtarge2019corpora}
Jared Lichtarge, Chris Alberti, Shankar Kumar, Noam Shazeer, Niki Parmar, and
  Simon Tong.
\newblock Corpora generation for grammatical error correction.
\newblock In \emph{Proceedings of the 2019 Conference of the North American
  Chapter of the Association for Computational Linguistics: Human Language
  Technologies, Volume 1 (Long and Short Papers)}, pp.\  3291--3301, 2019.

\bibitem[Maher(2018)]{Wikipedia_mirror_2018}
Katherine Maher.
\newblock Wikipedia is a mirror of the world’s gender biases, 2018.
\newblock URL
  \url{https://wikimediafoundation.org/news/2018/10/18/wikipedia-mirror-world-gender-biases/}.

\bibitem[Mako~Hill et~al.(2019)Mako~Hill, Champion, Shaw, Foote, McDonald,
  Forte, Narayan, and Greenstadt]{valueipediting}
Benjamin Mako~Hill, Kaylea Champion, Aaron Shaw, Jeremy Foote, Nora McDonald,
  Andrea Forte, Sneha Narayan, and Rachel Greenstadt.
\newblock Value of ip editing, 2019.
\newblock URL
  \url{https://meta.wikimedia.org/wiki/Research:Value_of_IP_Editing}.

\bibitem[Merity et~al.(2016)Merity, Xiong, Bradbury, and
  Socher]{merity2016pointer}
Stephen Merity, Caiming Xiong, James Bradbury, and Richard Socher.
\newblock Pointer sentinel mixture models.
\newblock \emph{arXiv preprint arXiv:1609.07843}, 2016.

\bibitem[Miquel-Rib{\'e} \& Laniado(2019)Miquel-Rib{\'e} and
  Laniado]{miquel2019wikipedia}
Marc Miquel-Rib{\'e} and David Laniado.
\newblock Wikipedia cultural diversity dataset: A complete cartography for 300
  language editions.
\newblock In \emph{Proceedings of the International AAAI Conference on Web and
  Social Media}, volume~13, pp.\  620--629, 2019.

\bibitem[Miquel-Rib{\'e} \& Laniado(2020)Miquel-Rib{\'e} and
  Laniado]{miquel2020wikipedia}
Marc Miquel-Rib{\'e} and David Laniado.
\newblock The wikipedia diversity observatory: A project to identify and bridge
  content gaps in wikipedia.
\newblock In \emph{Proceedings of the 16th International Symposium on Open
  Collaboration}, pp.\  1--4, 2020.

\bibitem[Mitrevski et~al.(2020)Mitrevski, Piccardi, and
  West]{mitrevski2020wikihist}
Blagoj Mitrevski, Tiziano Piccardi, and Robert West.
\newblock Wikihist.html: English wikipedia's full revision history in html
  format.
\newblock In \emph{Proceedings of the International AAAI Conference on Web and
  Social Media}, volume~14, pp.\  878--884, 2020.

\bibitem[Morgan et~al.(2013)Morgan, Gilbert, McDonald, and
  Zachry]{morgan2013project}
Jonathan~T Morgan, Michael Gilbert, David~W McDonald, and Mark Zachry.
\newblock Project talk: Coordination work and group membership in wikiprojects.
\newblock In \emph{Proceedings of the 9th International Symposium on Open
  Collaboration}, pp.\  1--10, 2013.

\bibitem[Nekoto et~al.(2020)Nekoto, Marivate, Matsila, Fasubaa, Fagbohungbe,
  Akinola, Muhammad, Kabenamualu, Osei, Sackey,
  et~al.]{nekoto2020participatory}
Wilhelmina Nekoto, Vukosi Marivate, Tshinondiwa Matsila, Timi Fasubaa, Taiwo
  Fagbohungbe, Solomon~Oluwole Akinola, Shamsuddeen Muhammad, Salomon~Kabongo
  Kabenamualu, Salomey Osei, Freshia Sackey, et~al.
\newblock Participatory research for low-resourced machine translation: A case
  study in african languages.
\newblock In \emph{Findings of the Association for Computational Linguistics:
  EMNLP 2020}, pp.\  2144--2160, 2020.

\bibitem[Ozurumba(2021)]{Ozurumba_2021}
Uzoma Ozurumba.
\newblock Content translation tool helps create one million wikipedia articles,
  Nov 2021.
\newblock URL
  \url{https://diff.wikimedia.org/2021/11/16/content-translation-tool-helps-create-one-million-wikipedia-articles/}.

\bibitem[Paullada et~al.(2021)Paullada, Raji, Bender, Denton, and
  Hanna]{paullada2021data}
Amandalynne Paullada, Inioluwa~Deborah Raji, Emily~M Bender, Emily Denton, and
  Alex Hanna.
\newblock Data and its (dis) contents: A survey of dataset development and use
  in machine learning research.
\newblock \emph{Patterns}, 2\penalty0 (11):\penalty0 100336, 2021.

\bibitem[Piccardi \& West(2021)Piccardi and West]{piccardi2021crosslingual}
Tiziano Piccardi and Robert West.
\newblock Crosslingual topic modeling with wikipda.
\newblock In \emph{Proceedings of the Web Conference 2021}, pp.\  3032--3041,
  2021.

\bibitem[Radford et~al.(2019)Radford, Wu, Child, Luan, Amodei, Sutskever,
  et~al.]{radford2019language}
Alec Radford, Jeffrey Wu, Rewon Child, David Luan, Dario Amodei, Ilya
  Sutskever, et~al.
\newblock Language models are unsupervised multitask learners.
\newblock \emph{OpenAI blog}, 1\penalty0 (8):\penalty0 9, 2019.

\bibitem[Rajpurkar et~al.(2016)Rajpurkar, Zhang, Lopyrev, and
  Liang]{rajpurkar2016squad}
Pranav Rajpurkar, Jian Zhang, Konstantin Lopyrev, and Percy Liang.
\newblock Squad: 100,000+ questions for machine comprehension of text.
\newblock In \emph{Proceedings of the 2016 Conference on Empirical Methods in
  Natural Language Processing}, pp.\  2383--2392, 2016.

\bibitem[Redi et~al.(2020)Redi, Gerlach, Johnson, Morgan, and
  Zia]{redi2020taxonomy}
Miriam Redi, Martin Gerlach, Isaac Johnson, Jonathan Morgan, and Leila Zia.
\newblock A taxonomy of knowledge gaps for wikimedia projects (second draft).
\newblock \emph{arXiv preprint arXiv:2008.12314}, 2020.

\bibitem[Sambasivan et~al.(2021)Sambasivan, Kapania, Highfill, Akrong,
  Paritosh, and Aroyo]{sambasivan2021everyone}
Nithya Sambasivan, Shivani Kapania, Hannah Highfill, Diana Akrong, Praveen
  Paritosh, and Lora~M Aroyo.
\newblock “everyone wants to do the model work, not the data work”: Data
  cascades in high-stakes ai.
\newblock In \emph{Proceedings of the 2021 CHI Conference on Human Factors in
  Computing Systems}, pp.\  1--15, 2021.

\bibitem[Sato(2021)]{Sato_2021}
Yumiko Sato.
\newblock Non-english editions of wikipedia have a misinformation problem.
\newblock \emph{Slate}, Mar 2021.
\newblock ISSN 1091-2339.
\newblock URL
  \url{https://slate.com/technology/2021/03/japanese-wikipedia-misinformation-non-english-editions.html}.

\bibitem[Schuhmann et~al.(2021)Schuhmann, Vencu, Beaumont, Kaczmarczyk, Mullis,
  Katta, Coombes, Jitsev, and Komatsuzaki]{schuhmann2021laion}
Christoph Schuhmann, Richard Vencu, Romain Beaumont, Robert Kaczmarczyk,
  Clayton Mullis, Aarush Katta, Theo Coombes, Jenia Jitsev, and Aran
  Komatsuzaki.
\newblock Laion-400m: Open dataset of clip-filtered 400 million image-text
  pairs.
\newblock \emph{arXiv preprint arXiv:2111.02114}, 2021.

\bibitem[Schwenk et~al.(2021)Schwenk, Chaudhary, Sun, Gong, and
  Guzm{\'a}n]{schwenk2021wikimatrix}
Holger Schwenk, Vishrav Chaudhary, Shuo Sun, Hongyu Gong, and Francisco
  Guzm{\'a}n.
\newblock Wikimatrix: Mining 135m parallel sentences in 1620 language pairs
  from wikipedia.
\newblock In \emph{Proceedings of the 16th Conference of the European Chapter
  of the Association for Computational Linguistics: Main Volume}, pp.\
  1351--1361, 2021.

\bibitem[Sen et~al.(2015)Sen, Ford, Musicant, Graham, Keyes, and
  Hecht]{sen2015barriers}
Shilad~W Sen, Heather Ford, David~R Musicant, Mark Graham, OS~Keyes, and Brent
  Hecht.
\newblock Barriers to the localness of volunteered geographic information.
\newblock In \emph{Proceedings of the 33rd Annual ACM Conference on Human
  Factors in Computing Systems}, pp.\  197--206, 2015.

\bibitem[Shi et~al.(2019)Shi, Teplitskiy, Duede, and Evans]{shi2019wisdom}
Feng Shi, Misha Teplitskiy, Eamon Duede, and James~A Evans.
\newblock The wisdom of polarized crowds.
\newblock \emph{Nature human behaviour}, 3\penalty0 (4):\penalty0 329--336,
  2019.

\bibitem[Smith et~al.(2022)Smith, Patwary, Norick, LeGresley, Rajbhandari,
  Casper, Liu, Prabhumoye, Zerveas, Korthikanti, et~al.]{smith2022using}
Shaden Smith, Mostofa Patwary, Brandon Norick, Patrick LeGresley, Samyam
  Rajbhandari, Jared Casper, Zhun Liu, Shrimai Prabhumoye, George Zerveas,
  Vijay Korthikanti, et~al.
\newblock Using deepspeed and megatron to train megatron-turing nlg 530b, a
  large-scale generative language model.
\newblock \emph{arXiv preprint arXiv:2201.11990}, 2022.

\bibitem[Srinivasan et~al.(2021)Srinivasan, Raman, Chen, Bendersky, and
  Najork]{srinivasan2021wit}
Krishna Srinivasan, Karthik Raman, Jiecao Chen, Michael Bendersky, and Marc
  Najork.
\newblock Wit: Wikipedia-based image text dataset for multimodal multilingual
  machine learning.
\newblock In \emph{Proceedings of the 44th International ACM SIGIR Conference
  on Research and Development in Information Retrieval}, pp.\  2443--2449,
  2021.

\bibitem[TeBlunthuis et~al.(2018)TeBlunthuis, Shaw, and
  Hill]{teblunthuis2018revisiting}
Nathan TeBlunthuis, Aaron Shaw, and Benjamin~Mako Hill.
\newblock Revisiting" the rise and decline" in a population of peer production
  projects.
\newblock In \emph{Proceedings of the 2018 CHI Conference on Human Factors in
  Computing Systems}, pp.\  1--7, 2018.

\bibitem[Voss(2005)]{voss2005measuring}
Jakob Voss.
\newblock Measuring wikipedia.
\newblock 2005.

\bibitem[Wang et~al.(2021)Wang, Tei, and Kohli]{ptwiki2021}
Jennifer~Wang Wang, Sandister Tei, and Niharika Kohli.
\newblock Impact report of turning off ip editing on portuguese wikipedia,
  2021.
\newblock URL
  \url{https://meta.wikimedia.org/wiki/IP_Editing:_Privacy_Enhancement_and_Abuse_Mitigation/Impact_report_for_IP_Editing_Restriction_Study_on_Portuguese_Wikipedia}.

\bibitem[Warncke-Wang et~al.(2012)Warncke-Wang, Uduwage, Dong, and
  Riedl]{warncke2012search}
Morten Warncke-Wang, Anuradha Uduwage, Zhenhua Dong, and John Riedl.
\newblock In search of the ur-wikipedia: universality, similarity, and
  translation in the wikipedia inter-language link network.
\newblock In \emph{Proceedings of the Eighth Annual International Symposium on
  Wikis and Open Collaboration}, pp.\  1--10, 2012.

\bibitem[Wikimedia~Foundation(2021{\natexlab{a}})]{croatianreport}
Trust \& Safety Disinformation~Team Wikimedia~Foundation.
\newblock The case of croatian wikipedia: Encyclopaedia of knowledge or
  encyclopaedia for the nation?, 2021{\natexlab{a}}.
\newblock URL
  \url{https://meta.wikimedia.org/wiki/Croatian_Wikipedia_Disinformation_Assessment-2021}.

\bibitem[Wikimedia~Foundation(2021{\natexlab{b}})]{ucocresearch}
Trust \& Safety Disinformation~Team Wikimedia~Foundation.
\newblock Universal code of conduct/research - wikipedia, 2021{\natexlab{b}}.
\newblock URL
  \url{https://meta.wikimedia.org/wiki/Universal_Code_of_Conduct/Research_-_Wikipedia\#Policy_links,_description,_and_uniqueness_rating}.

\bibitem[Wu \& Dredze(2020)Wu and Dredze]{wu2020all}
Shijie Wu and Mark Dredze.
\newblock Are all languages created equal in multilingual bert?
\newblock In \emph{Proceedings of the 5th Workshop on Representation Learning
  for NLP}, pp.\  120--130, 2020.

\end{thebibliography}
\bibliographystyle{iclr2022_conference}

\end{document}